\documentclass[sigconf]{acmart}

\usepackage{booktabs} 
\usepackage{hhline}
\usepackage{graphicx}

\setcopyright{rightsretained}

\acmDOI{10.475/123_4}

\acmISBN{123-4567-24-567/08/06}

\acmConference[SocVis17]{SocVis technical report}{June 2017}{}
\acmYear{2017}
\copyrightyear{2017}

\acmPrice{00.00}

\begin{document}
\title{Towards a Recommender System for Undergraduate Research}

\author{Felipe del Rio}
\affiliation{%
  \institution{Pontificia Universidad Catolica de Chile}
  \city{Santiago}
  \state{Chile}
}
\email{fidelrio@uc.cl}

\author{Denis Parra}
\affiliation{%
  \institution{Pontificia Universidad Catolica de Chile}
  \city{Santiago}
  \state{Chile}
}
\email{dparra@ing.puc.cl}

\author{Jovan Kuzmicic}
\affiliation{%
  \institution{Pontificia Universidad Catolica de Chile}
  \city{Santiago}
  \state{Chile}
}
\email{jpkuzmic@ing.puc.cl}

\author{Erick Svec}
\affiliation{%
  \institution{Pontificia Universidad Catolica de Chile}
  \city{Santiago}
  \state{Chile}
}
\email{evsvec@ing.puc.cl}

\renewcommand{\shortauthors}{Author et al.}

\begin{abstract}
Several studies indicate that attracting students to research careers requires to engage them from early undergraduate years. Following this paradigm, our Engineering School has developed an undergraduate research program that allows students to enroll in research in exchange for course credits. Moreover, we developed a web portal to inform students about the program and the opportunities, but participation remains lower than expected.
In order to promote student engagement, we attempt to build a personalized recommender system of research opportunities to undergraduates.
With this goal in mind we investigate two tasks. First, one that identifies students that are more willing to participate on this kind of program. A second task is generating  a personalized list of recommendations of research opportunities for each student.
To evaluate our approach, we perform a simulated prediction experiment with data from our School, which has more than 4,000 active undergraduate students nowadays. Our results indicate that there is a big potential to create a personalized recommender system for this purpose.
Our results can be used as a baseline for colleges seeking strategies to encourage research activities within undergraduate students.
\end{abstract}

\keywords{Recommender Systems, Undergraduate Research}

\maketitle

\section{Introduction}

In a globalized world, academic institutions are compelled to offer rich learning experiences to their students, with a complex curriculum that include extra academic activities \cite{bauer2003alumni}. In order to address this issue, our School of Engineering\footnote{institution not disclosed to maintain blind revision requirement} established an undergraduate research program in 2011, known as IPre (in Spanish \textit{Investigaci\'on en Pregrado}), which allows students to receive course credits when joining a research project with faculty advice. The mission of the IPre program is to contribute to the academic and professional development of engineering undergraduates by enhancing their research skills \cite{harsh2011undergraduate}.

\textbf{Context and Problem}. Nowadays, the IPre program has an offer-demand system focused on student-faculty interaction on a web platform. Herein, professors offer \textit{Research Opportunities} to a general board where students can browse and apply to available projects. In this way, students have access to research topics that are new to them and work in different attractive areas. Although this platform promotes exchange of ideas, student engagement in undergraduate research programs faces major challenges \cite{merkel2001undergraduate}, and IPre is not an exception. In order to promote these programs, recent literature has aimed to identify undergraduates' motivation with research activities \cite{douglass2013undergraduate,zimbardi2014embedding}. In this line, we have detected lack of knowledge about the IPre program and the available research opportunities as a major factor, thus we herein propose a personalized approach to enroll students in undergraduate research.

\textbf{Objective and Tasks}. In order to address the challenge of promoting student engagement in our undergraduate research, and considering the success of personalization for increasing user engagement in several areas and communities, we decided to explore the potential of a recommender systems. In this work we study the feasibility of such system studying two tasks, using data collected from the current online IPre system over the last five years: (i) Identifying Students who would be likely to participate in the undergraduate research program, and (ii) recommending relevant research opportunities to undergraduate Engineering students.

\textbf{Results and Contributions}. Our results indicate that it is possible to identify which students will be more likely to participate, with a precision up to 72.7\%. Moreover, the task of recommending is indeed more challenging. We compared several methods and parameters and we were able to obtain a model which close to MAP=0.2, but it requires further research to get to a more accurate recommendation approach. Nonetheless, these results set an appropriate baseline to improve further our current IPre system.

\section{Dataset \& Features}

We used a dataset from the IPre program over 2012-2016 period, representing applications of students to undergraduate research opportunities. The dataset comprises user profiles of $10,546$ undergraduate students of the Engineering School, among them $1,134$ students applied to $1,017$ available research opportunities. Students could apply to more than one opportunity, so we recorded $1,624$ applications in total, having 81.4\% of the applications accepted.

{\it Task 1} was about predicting whether student $u_i$ applied to research opportunities or not (1:applied, 0:did not apply). In this task we compared three feature sets: (a) {\it Base}: semesters enrolled, number of credits approved, (b) {\it Base + ipre:} features in (a) plus a boolean indicating previous applications to IPRE, and (c) {\it Base + ipre + gpa:} features in (b) plus GPA.

For {\it Task 2}--predicting which research opportunities the students applied-- we made recommendation as a classification task, i.e., predict whether student $u_i$ would apply to a research opportunity $o_j$ (1:positive, 0:negative). We used three feature sets: (a) {\it Base}: cosine similarity between research opportunity abstract and descriptions of courses approved, (b) {\it Base + ht}: features in (a) plus a boolean indicating that the student was taught by the faculty offering the opportunity, and (c) {\it Base + ht + dept} features in (b) plus the percentage of courses approved taught by the same department as the faculty offering the opportunity (e.g. computer science).

\def\arraystretch{0.8}%
\begin{table}[t!]
\caption{Task 1, predict if student applies to opportunities.}
\label{table:is-ipre-student}
    \begin{center}
    \scalebox{1.0}{
        \begin{tabular}{ l  c  c  c }
        \toprule
                        & Accuracy          & Precision         & F-1 Score     \\ 
\midrule
        Baseline        & 10.9\%            & 10.9\%            & 0.20          \\ 
        LogReg          & 91.2\%            & 62.4\%            & \textbf{0.55} \\ 
        GBT             & \textbf{92.0\%}   & \textbf{72.7\%}   & 0.54          \\ 
        SVM             & 90.1\%            & 67.4\%            & 0.28          \\ 
\midrule
        Base (GBT)            & 89.1\%            & 25.0\%            & 0.01          \\ 
        Base+ipre  (GBT)      & \textbf{92.1\%}   & 71.7\%            & \textbf{0,55} \\ 
        Base+ipre+gpa  (GBT)  & 92.0\%            & \textbf{72.7\%}   & 0.54          \\ 
    \bottomrule
        \end{tabular}
        }
    \end{center}
\end{table}

\section{Evaluation Methodology \& Results}

All data before 2014 is used for training and everything afterwards for testing. In both tasks we test a baseline classifier, logistic regression (LogReg), gradient boosted trees (GBT) and support vector machines (SVM).
For \textbf{task 1}, predicting whether the user applies to opportunities or not, the dataset is highly unbalanced since 89.7\% of the students do not apply to opportunities. We  measure classifier performance with accuracy, precision and F-1 score. As a baseline we use a model that predicts the most common class.

For \textbf{task 2}, predicting which opportunities a student actually applied to, we classify several opportunities for each student and we rank them based on their prediction score. Then, we used the ranking metric Mean Average Precision (MAP) \cite{parra2013recommender} to evaluate the performance. The baseline method consisted on generating a random list of recommendations. In this task, we analyzed: recommendation list size ($k$), feature sets and algorithm used.

\section{Results}

\textbf{Task 1}: {\it Predict is student applies to opportunities}. Table \ref{table:is-ipre-student} shows the results in two groups: (a) comparing methods (using all features), and (b) comparing features (using the best method). Here we see that all methods (LogReg, GBT and SVM) outperform the baseline in all metrics. The best methods though are GBT (accuracy=92\%, precision=72.7\%, F-1=0.54) and LogReg (accuracy=91.2\%, precision=62.4\%, F-1=0.55). This result is very high considering the class imbalance. In terms of feature sets, the baseline (semesters enrolled and number of credits approved) is boosted specially by considering if the student previously applied to an IPre opportunity in the past; i.e., most likely will apply again.

\begin{figure}[t!]
\centering
  \includegraphics[scale=0.33]{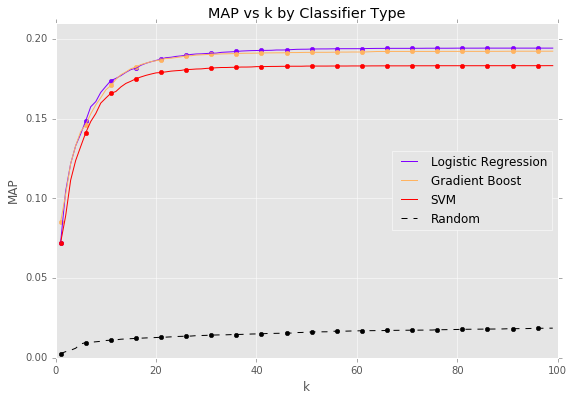}
\caption{Task2 MAP by classifier using all features.}
  \label{fig:map-by-type}
\end{figure}

\begin{figure}[t!]
\centering
\includegraphics[scale=0.33]{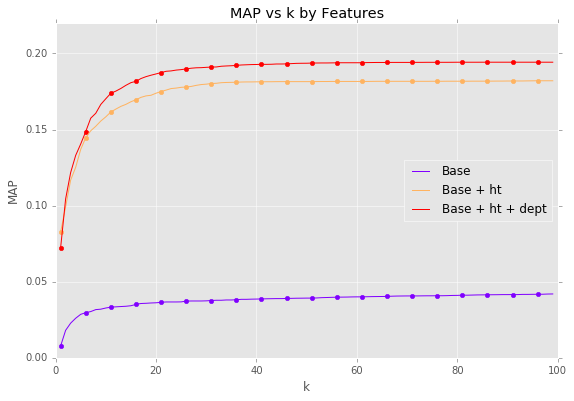}
  \caption{Task 2 MAP by feature sets using LR.}
  \label{fig:map-by-feat}
\end{figure}

\textbf{Task 2}: {\it Recommending research opportunities}.
We analyze this task in two stages. First, using all the features we compare methods, as seen in Figure \ref{fig:map-by-type}. We found that all methods outperform a random baseline, but LogReg and GBT perform the best, getting to a MAP up to $0.20$. Our top method scored 14.6 times higher that the baseline for $k=20$ and closer to 10 times on a longer recommendation list. Then, using LogReg method, we study different features set as seen in Figure \ref{fig:map-by-feat}. We observe that knowing if the student had a class with the professor offering the research opportunity increases significantly the prediction compared to only matching content description of courses and research opportunity. A smaller yet important boost on the recommendation is also given by matching department information in the model


\section{Conclusion}

In this work we showed feasibility of: (a) identifying students prone to apply to research opportunities, and (b) recommending research opportunities for undergraduate students. There is still room for improvement by adding new features and other recommendation approaches (such as factorization machines or neural networks). We are currently conducting a user study to verify the generalizability of our results. We expect to serve as a baseline for institutions implementing these features in their academic systems.

\bibliographystyle{ACM-Reference-Format}
\bibliography{sigproc}

\end{document}